\documentclass[journal]{IEEEtran}
\usepackage{amsfonts,subfigure,multicol,color,verbatim,
graphicx,cite,epsfig,amssymb,amsmath,cases,bm,algorithm,
algorithmic,xcolor,multirow}

\newtheorem{prob}{Problem}

\begin{document}
\title{Energy-Efficient Resource Assignment and Power Allocation in Heterogeneous Cloud Radio Access Networks}

\author{\IEEEauthorblockN{Mugen~Peng,~\IEEEmembership{Senior Member,~IEEE}, Kecheng~Zhang, Jiamo~Jiang, Jiaheng~Wang,~\IEEEmembership{Member,~IEEE}, and Wenbo~Wang,~\IEEEmembership{Member,~IEEE}}

\thanks{Copyright (c) 2015 IEEE. Personal use of this material is permitted. However, permission to use this material for any other purposes must be obtained from the IEEE by sending a request to pubs-permissions@ieee.org. }
\thanks{Mugen~Peng (e-mail: {\tt pmg@bupt.edu.cn}), Kecheng~Zhang (e-mail: {\tt buptzkc@163.com}), and Wenbo~Wang (e-mail: {\tt wbwang@bupt.edu.cn}) are with the Key Laboratory of Universal Wireless Communications for Ministry of Education, Beijing University of Posts and Telecommunications, Beijing, China. Jiamo~Jiang (e-mail: {\tt jiamo.jiang@gmail.com}) is with the Institute of Communication Standards Research, China Academy of Telecommunication Research of MIIT, Beijing, China. Jiaheng~Wang (e-mail: {\tt jhwang@seu.edu.cn}) is with the National Mobile Communications Research Laboratory, Southeast University, Nanjing, China.}
\thanks{The work of M. Peng, K. Zhang and W. Wang was supported in part by the National Natural Science Foundation
of China (Grant No. 61222103), the National Basic Research Program
of China (973 Program) (Grant No. 2013CB336600, 2012CB316005), the
State Major Science and Technology Special Projects (Grant No.
2013ZX03001001), the Beijing Natural Science Foundation (Grant No.
4131003), and the Specialized Research Fund for the Doctoral Program
of Higher Education (SRFDP) (Grant No.20120005140002). The work of
J. Wang was supported by the National Natural Science Foundation of
China under Grant 61201174, by the Natural Science Foundation of
Jiangsu Province under Grant BK2012325, and by the Fundamental
Research Funds for the Central Universities.}}

\maketitle

\begin{abstract}

Taking full advantages of both heterogeneous networks (HetNets) and
cloud access radio access networks (C-RANs), heterogeneous cloud
radio access networks (H-CRANs) are presented to enhance both the
spectral and energy efficiencies, where remote radio heads (RRHs)
are mainly used to provide high data rates for users with high
quality of service (QoS) requirements, while the high power node
(HPN) is deployed to guarantee the seamless coverage and serve users
with low QoS requirements. To mitigate the inter-tier interference
and improve EE performances in H-CRANs, characterizing user
association with RRH/HPN is considered in this paper, and the
traditional soft fractional frequency reuse (S-FFR) is enhanced.
Based on the RRH/HPN association constraint and the enhanced S-FFR,
an energy-efficient optimization problem with the resource
assignment and power allocation for the orthogonal frequency
division multiple access (OFDMA) based H-CRANs is formulated as a
non-convex objective function. To deal with the non-convexity, an
equivalent convex feasibility problem is reformulated, and
closed-form expressions for the energy-efficient resource allocation
solution to jointly allocate the resource block and transmit power
are derived by the Lagrange dual decomposition method. Simulation
results confirm that the H-CRAN architecture and the corresponding
resource allocation solution can enhance the energy efficiency
significantly.

\end{abstract}

\section{Introduction}

Cloud radio access networks (C-RANs) are by now recognized to
curtail the capital and operating expenditures, as well as to
provide a high transmission bit rate with fantastic energy
efficiency (EE)
performances\textcolor[rgb]{1.00,0.00,0.00}{\cite{bib:CRAN,bib:TDSCDMA}}.
The remote radio heads (RRHs) operate as soft relay by compressing
and forwarding the received signals from the mobile user equipment
(UE) to the centralized base band unit (BBU) pool through the
wire/wireless fronthaul links. To highlight the advantages of C-RAN,
the joint decompression and decoding schemes are executed in the BBU
pool\textcolor[rgb]{1.00,0.00,0.00}{\cite{bib:Park}}. However, the
non-ideal fronthaul with limited capacity and long time delay
degrades performances of C-RANs. Furthermore, it is critical to
decouple the control and user planes in C-RANs, and RRHs are
efficient to provide high capacity without considering functions of
control planes. How to alleviate the negative influence of the
constrained fronthaul on EE performances, and how to broadcast the
control signallings to UEs without RRHs are still not
straightforward in
C-RANs\textcolor[rgb]{1.00,0.00,0.00}{\cite{bib:Park2}}.

Meanwhile, high power nodes (HPNs) (e.g., macro or micro base
stations) existing in heterogeneous networks (HetNets) are still
critical to guarantee the backward compatibility with the
traditional cellular networks and support the seamless coverage
since low power nodes (LPNs) are mainly deployed to provide high bit
rates in special
zones\textcolor[rgb]{1.00,0.00,0.00}{\cite{bib:HetNet}}. Under help
of HPNs, the unnecessary handover can be avoided and the synchronous
constraints among LPNs can be alleviated. Accurately, although the
HetNet is a good alternative to improve both coverage and capacity
simultaneously, there are two remarkable challenges to block its
commercial applications: \romannumeral1). The coordinated
multi-point transmission and reception (CoMP) needs a huge number of
signallings in backhaul links to mitigate the inter-tier
interferences between HPNs and LPNs, while the backhaul capacity is
often constrained; \romannumeral2). The ultra dense LPNs improve
capacity with the cost of consuming too much energy, which results
in low EE performances.

\begin{figure}
\centering  \vspace*{0pt}
\includegraphics[scale=0.35]{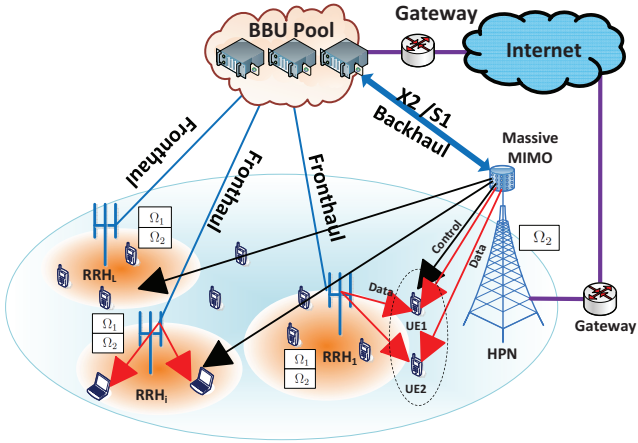}
\setlength{\belowcaptionskip}{-100pt} \caption{\textbf{System
architecture of the proposed H-CRANs}} \label{HRAN}\vspace*{-10pt}
\end{figure}

To overcome these aforementioned challenges in both C-RANs and
HetNets, a new architecture and technology known as the
heterogeneous cloud radio access network (H-CRAN) is presented as a
promising paradigm for future heterogeneous converged
networks\textcolor[rgb]{1.00,0.00,0.00}{\cite{MPeng_HCRAN}}. The
H-CRAN architecture shown in Fig. \ref{HRAN} takes full advantages
of both C-RANs and HetNets, where RRHs with low energy consumptions
are cooperated with each other in the BBU pool to achieve high
cooperative gains. The BBU pool is interfaced to HPNs for
coordinating the inter-tier interferences between RRHs and HPNs.
Only the front radio frequency (RF) and simple symbol processing
functionalities are configured in RRHs, while the other important
baseband physical processing and procedures of the upper layers are
executed in the BBU pool. By contrast, the entire communication
functionalities from the physical to network layers are implemented
in HPNs. The data and control interfaces between the BBU pool and
HPNs are S1 and X2, respectively, which are inherited from
definitions of the $3^{rd}$ generation partnership project (3GPP)
standards. Compared with the traditional C-RAN architecture
in\textcolor[rgb]{1.00,0.00,0.00}{\cite{bib:CRAN}}, H-CRANs
alleviate the fronthaul constraints between RRHs and the BBU pool
through incorporating HPNs. The control and broadcast
functionalities are shifted from RRHs to HPNs, which alleviates
capacity and time delay constraints on the fronthaul and supports
the burst traffic efficiently. The adaptive signalling/control
mechanism between connection-oriented and connectionless is
supported in H-CRANs, which can achieve significant overhead savings
in the radio connection/release by moving away from a pure
connection-oriented mechanism.

In H-CRANs, the RRH/HPN association strategy is critical for
improving EE performances, and main differences from the traditional
cell association techniques are twofold. First, the transmit power
of RRHs and HPNs is significantly different. Second, the inter-RRH
interferences can be jointly coordinated through the centralized
cooperative processing in the BBU pool, while the inter-tier
interferences between HPNs and RRHs are severe and difficult to
mitigate. Consequently, it is not always efficient for UEs to be
associated with neighbor RRHs/HPNs via the strongest receiving power
mechanism\textcolor[rgb]{1.00,0.00,0.00}{\cite{bib:CRAN_CWL}}. As
shown in Fig. \ref{HRAN}, though obtaining the same receiving power
from RRH$_1$ and HPN, both UE1 and UE2 prefer to associate with
RRH$_1$ because lower transmit power is needed and more radio
resources are allocated from RRH$_1$ than those from the HPN.
Meanwhile, the association with RRHs can decrease energy
consumptions by saving the massive use of air condition.

Based on the aforementioned RRH/HPN association characteristics, the
joint optimization solution for resource block (RB) assignment and
power allocation to maximize EE performances in the orthogonal
frequency division multiple access (OFDMA) based H-CRANs is
researched in this paper.

\subsection{Related Work}

OFDMA is a promising multi-access technique for exploiting channel
variations in both frequency and time domains to provide high data
rates in the fourth generation (4G) and beyond cellular networks. To
be backward compatible with 4G systems, the OFDMA is adopted in
H-CRANs by assigning different RBs to different UEs. Recently, the
radio resource allocation (RA) to maximize the spectral efficiency
(SE) or meet diverse qualify-of-service (QoS) requirements in OFDMA
systems has attracted considerable
attention\textcolor[rgb]{1.00,0.00,0.00}{\cite{b3} -
\cite{chuanghuang}}. The relay selection problem in a network with
multiple UEs and multiple amplify-and-forward relays is investigated
in\textcolor[rgb]{1.00,0.00,0.00}{\cite{b3}}, where an optimal relay
selection scheme whose complexity is quadratic in the number of UEs
and relays is presented. The authors
in\textcolor[rgb]{1.00,0.00,0.00}{\cite{b4}} propose an asymptotical
resource allocation algorithm via leveraging the cognitive radio
(CR) technique in open access OFDMA femtocell networks. The resource
optimization for spectrum sharing with interference control in CR
systems is researched in\textcolor[rgb]{1.00,0.00,0.00}{\cite{b5}},
where the achievable transmission rate of the secondary user over
Rayleigh channels subject to a peak power constraint at the
secondary transmitter and an average interference power constraint
at the primary receiver is maximized.
In\textcolor[rgb]{1.00,0.00,0.00}{\cite{chuanghuang}}, the optimal
power allocation for minimizing the outage probability in
point-to-point fading channels with the energy-harvesting
constraints is investigated.

The EE performance metric has become a new design goal due to the
sharp increase of the carbon emission and operating cost of wireless
communication systems\textcolor[rgb]{1.00,0.00,0.00}{\cite{b6}}. The
EE-oriented radio resource allocation has been studied in various
networks\textcolor[rgb]{1.00,0.00,0.00}{\cite{b7} - \cite{b10}}.
In\textcolor[rgb]{1.00,0.00,0.00}{\cite{b7}}, the distributed power
allocation for multi-cell OFDMA networks taking both energy
efficiency and inter-cell interference mitigation into account is
investigated, where a bi-objective problem is formulated based on
the multi-objective optimization theory. To maximize the average EE
performance of multiple UEs each with a transceiver of constant
circuit power, the power allocation, RB allocation and relay
selection are jointly optimized
in\textcolor[rgb]{1.00,0.00,0.00}{\cite{b8}}. The active number of
sub-carriers and the number of bits allocated to each RB at the
source nodes are optimized to maximize EE performances
in\textcolor[rgb]{1.00,0.00,0.00}{\cite{b9}}, where the optimal
solution turns out to be a bidirectional water-filling bit
allocation to minimize the overall transmit power. To maximize EE
performances under constraints of total transmit power and
interference in CR systems, an optimal power allocation algorithm
using equivalent conversion is proposed
in\textcolor[rgb]{1.00,0.00,0.00}{\cite{b10}}.

Intuitively, the inter-cell or inter-tier interference mitigation is
the key to improve both SE and EE performances. Some advanced
algorithms in HetNets, such as cell association and fractional
frequency reuse (FFR), have been proposed
in\textcolor[rgb]{1.00,0.00,0.00}{\cite{b11}}
and\textcolor[rgb]{1.00,0.00,0.00}{\cite{b12}}, respectively.
In\textcolor[rgb]{1.00,0.00,0.00}{\cite{WeiYu}}, a network utility
maximization formulation with a proportional fairness objective is
presented, where prices are updated in the dual domain via
coordinate descent. In\textcolor[rgb]{1.00,0.00,0.00}{\cite{b13}},
energy efficient cellular networks through the employment of base
station with sleep mode strategies as well as small cells are
researched, and the corresponding tradeoff issue is discussed as
well.

To the best of our knowledge, there are lack of solutions to
maximize the EE performance in H-CRANs. Particularly, the RRH/HPN
association strategy should be enhanced from the traditional
strongest received power strategy. Furthermore, the radio resource
allocation to achieve an optimal EE performance in H-CRANs is still
not straightforward. To deal with these problems, the joint
optimization solution with the RB assignment and power allocation
subject to the RRH/HPN association and interference mitigation
should be investigated.

\subsection{Main Contributions}

The goal of this paper is to investigate the joint optimization
problem with the RB assignment and power allocation subject to the
RRH/HPN association and inter-tier interference mitigation to
maximize EE performances in the OFDMA based H-CRAN system. The EE
performance optimization is highly challenging because the
energy-efficient resource allocation in H-CRANs is a non-convex
objective problem. Different from the published radio resource
optimization in HetNets and C-RANs, the characteristics of H-CRANs
should be highlighted and modeled. To simplify the coordinated
scheduling between RRHs and the HPN, an enhanced soft fractional
frequency reuse (S-FFR) scheme is presented to improve performances
of cell-center-zone UEs served by RRHs with individual spectrum
frequency resources. The contributions of this paper can be
summarized as follows:

\begin{itemize}
  \item To overcome challenges in HetNets and C-RANs, H-CRANs are presented
  as cost-efficient potential solutions to improve spectral and energy efficiencies, in which RRHs are mainly used to provide high
data rates for UEs with high QoS requirements in the hot spots,
while HPNs are deployed to guarantee seamless coverage for UEs with
low QoS requirements.
  \item To mitigate the inter-tier interference
between RHHs and HPNs, an enhanced S-FFR scheme is proposed, where
the total frequency band is divided into two parts. Only partial
spectral resources are shared by RRHs and HPNs, while the other is
solely occupied by RRHs. The exclusive RBs are allocated to UEs with
high rate-constrained QoS requirements, while the shared RBs are
allocated to UEs with low rate-constrained QoS requirements.
  \item An energy-efficient optimization problem with the
RB assignment and power allocation under constraints of the
inter-tier interference mitigation and RRH/HPN association is
formulated as a non-convex objective function. To deal with this
non-convexity, an equivalent convex feasibility problem is
reformulated, based on which an iterative algorithm consisting of
both outer and inner loop optimizations is proposed to achieve the
global optimal solution.
  \item We numerically evaluate EE performance gains of H-CRANs and the corresponding resource allocation optimization
solution. Simulation results demonstrate that EE performance gain of
the H-CRAN architecture over the traditional C-RAN/HetNet is
significant. The proposed iterative solution is converged, and its
EE performance gain over the baseline algorithms is impressive.
\end{itemize}

The reminder of this paper is organized as follows. In Section II,
we describe the system model of H-CRANs, the proposed enhanced
S-FFR, and the problem formulation. The optimization framework is
introduced in Section III. Section IV provides simulations to verify
the effectiveness of the proposed H-CRAN architecture and the
corresponding solutions. Finally, we conclude the paper in Section
V.

\section{System Model and Problem Formulation}

\begin{figure}
\centering  \vspace*{0pt}
\includegraphics[scale=0.31]{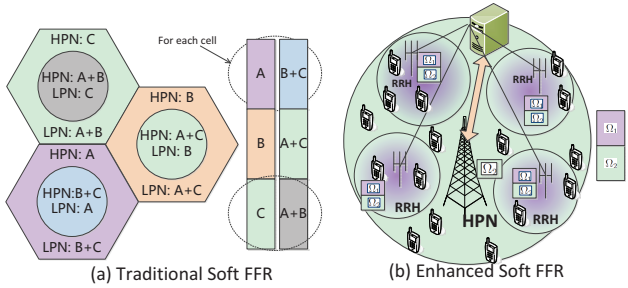}
\setlength{\belowcaptionskip}{-100pt} \caption{\textbf{Principle of
the proposed enhanced soft fractional frequency reuse scheme}}
\label{FFR}\vspace*{-10pt}
\end{figure}

The traditional S-FFR is considered as an efficient inter-cell and
inter-tier interference coordination technique, in which the service
area is partitioned into spatial subregions, and each subregion is
assigned with different frequency sub-bands. As shown in Fig.
\ref{FFR}(a), the cell-edge-zone UEs do not interfere with
cell-center-zone UEs, and the inter-cell interference can be
suppressed with an efficient channel allocation
method\textcolor[rgb]{1.00,0.00,0.00}{\cite{SFR}}. The HPN is mainly
used to deliver the control signallings and guarantee the seamless
coverage for UEs accessing the HPN (denoted by HUEs) with low QoS
requirements. By contrast, the QoS requirement for UEs accessing RRH
(denoted by RUEs) is often with a higher priority. Consequently, as
shown in Fig. \ref{FFR}(b), an enhanced S-FFR scheme is proposed to
mitigate the inter-tier interference between HPNs and RRHs, in which
only partial radio resources are allocated to both RUEs and HUEs
with low QoS requirements, and the remaining radio resources are
allocated to RUEs with high QoS requirements. In the proposed
enhanced S-FFR, RUEs with low QoS requirements share the same radio
resources with HUEs, which is absolutely different from that in the
traditional S-FFR. If the traditional S-FFR is utilized in H-CRANs,
only the cell-center-zone RUEs share the same radio resources with
HUEs, which decreases the SE performance significantly. Further, it
is challenging to judge whether UEs are located in the cell-edge or
cell-center zone for the traditional S-FFR.

These aforementioned problems are avoided in the proposed enhanced
S-FFR, where only the QoS requirement should be distinguished for
RUEs. On the one hand, to avoid the inter-tier interference, the
outband frequency is preferred to use according to standards of
HetNets in 3GPP\textcolor[rgb]{1.00,0.00,0.00}{\cite{3GPP_HetNet}},
which suggests that RBs for HPNs should be orthogonal with those for
RRHs. On the other hand, to save the occupied frequency bands, the
inband strategy is defined as well in
3GPP\textcolor[rgb]{1.00,0.00,0.00}{\cite{3GPP_HetNet}}, which
indicates that both RUEs and HUEs share the same RBs even though the
inter-tier interference is severe. To be completely compatible with
both inband and outband strategies in 3GPP, only two RB sets
${\Omega _1}$ and ${\Omega _2}$ are divided in this proposed
enhanced S-FFR scheme. Obviously, if locations of RUEs could be
known and the traffic volume in different zones are clearly
anticipated, more RB sets in ${\Omega _1}$ could be divided to
achieve higher performance gains. The division of two RB sets in the
proposed enhanced S-FFR is a good tradeoff between performance gains
and implementing complexity/flexibility.

The QoS requirement is treated as the minimum transmission rate in
this paper, which is also called the rate-constrained QoS
requirement. In this paper, the high and low rate-constrained QoS
requirements are denoted as $\eta _{\rm{R}}$ and $\eta _{\rm{ER}}$,
respectively. For simplicity, it is assumed that there are $N$ and
$M$ RUEs per RRH occupying the RB sets ${\Omega _1}$ and ${\Omega
_2}$, respectively. In the OFDMA based downlink H-CRANs, there are
total $K$ RBs (denoted as ${\Omega _T}$) with the bandwidth ${B_0}$.
These $K$ RBs in ${\Omega _T}$ are categorized as two types:
${\Omega _1}$ is only allocated to RUEs with high rate-constrained
QoS requirements, and ${\Omega _2}$ is allocated to RUEs and HUEs
with low rate-constrained QoS requirements. Since all signal
processing for different RRHs is executed on the BBU pool centrally,
the inter-RRH interferences can be coordinated and the same radio
resources can be shared amongst RRHs. Hence, the
channel-to-interference-plus-noise ratio (CINR) for the \emph{n}-th
RUE occupying the \emph{k}-th RB can be divided into two parts:
\begin{equation}\label{eqn:CINR1} {\sigma _{n,k}} = \left\{
{\begin{array}{*{20}{c}}
{d_n^Rh_{n,k}^R{\rm{ }}/{B_0}{N_0}{\kern 1pt} {\kern 1pt} {\kern 1pt} {\kern 1pt} {\kern 1pt} {\kern 1pt} {\kern 1pt} {\kern 1pt} {\kern 1pt} {\kern 1pt} {\kern 1pt} {\kern 1pt} {\kern 1pt} {\kern 1pt} {\kern 1pt} {\kern 1pt} {\kern 1pt} {\kern 1pt} {\kern 1pt} {\kern 1pt} {\kern 1pt} {\kern 1pt} {\kern 1pt} {\kern 1pt} {\kern 1pt} {\kern 1pt} {\kern 1pt} {\kern 1pt} {\kern 1pt} {\kern 1pt} {\kern 1pt} {\kern 1pt} {\kern 1pt} {\kern 1pt} {\kern 1pt} {\kern 1pt} {\kern 1pt} {\kern 1pt} {\kern 1pt} {\kern 1pt} {\kern 1pt} {\kern 1pt} {\kern 1pt} {\kern 1pt} {\kern 1pt} {\kern 1pt} {\kern 1pt}{\kern 1pt}{\kern 1pt}{\kern 1pt}{\kern 1pt}{\kern 1pt}{\kern 1pt},k \in {\Omega _{\rm{1}}}}\\
{d_n^Rh_{n,k}^R{\rm{ }}/(P_{}^{\rm{M}}d_n^{\rm{M}}h_{n,k}^{\rm{M}} +
{B_0}{N_0}){\kern 1pt} {\kern 1pt} {\kern 1pt} {\kern 1pt} {\kern
1pt} {\kern 1pt} {\kern 1pt} {\kern 1pt} {\kern 1pt} ,k \in {\Omega
_{{\rm{2}}}} {\kern
1pt} {\kern 1pt} {\kern 1pt} {\kern 1pt}
{\kern 1pt} {\kern 1pt} {\kern 1pt} {\kern 1pt} {\kern 1pt} {\kern
1pt} {\kern 1pt} {\kern 1pt} {\kern 1pt} {\kern 1pt} {\kern 1pt}
{\kern 1pt} {\kern 1pt} {\kern 1pt} {\kern 1pt} {\kern 1pt} {\kern
1pt} {\kern 1pt} {\kern 1pt} {\kern 1pt} }
\end{array}} \right.
\end{equation}
where $d_n^{{\rm{R}}}$ and $d_n^{{\rm{M}}}$ denote the path loss
from the served RRH and the reference HPN to RUE $n$, respectively.
$h_{n,k}^{{\rm{R}}}$ and $h_{n,k}^{{\rm{M}}}$ represent the channel
gain from the RRH and HPN to RUE $n$ on the $k$-th RB, respectively.
$P_{}^{\rm{M}} = \frac{P_{\rm{max}}^{\rm{M}}}{M}$ is the allowed
transmit power allocated on each RB in HPN and
$P_{\rm{max}}^{\rm{M}}$ denotes the maximum allowable transmit power
of HPN. $N_0$ denotes the estimated power spectrum density (PSD) of
both the sum of noise and weak inter-RRH interference (in dBm/Hz).

The sum data rate for each RRH can be expressed as:
\begin{equation}\label{eqn:rates}
C({\bf{a}},{\bf{p}}) = \sum\limits_{n = 1}^{N + M} {\sum\limits_{k =
1}^K {{a_{n,k}}{B_0}{{\log }_2}(1 + {\sigma _{n,k}}{p_{n,k}})} },
\end{equation}
where $n \in \{ 1,...,N\}$ denotes the RUE allocated to the RB set
${\Omega _1}$, and $n \in \{ N + 1,...,N + M\}$ denotes the RUE
allocated to the RB set ${\Omega _2}$. The $(N + M) \times K$
matrices ${\bf{a}} = {\left[ {{a_{n,k}}} \right]_{(N + M) \times
K}}$ and ${\bf{p}} = {\left[ {{p_{n,k}}} \right]_{(N + M) \times
K}}$ represent the feasible RB and power allocation policies,
respectively. ${a_{n,k}}$ is defined as the RB allocation indicator
which can only be 1 or 0, indicating whether the  $k$-th RB is
allocated to RUE $n$. ${p_{n,k}}$ denotes the transmit power
allocated to RUE $n$ on the $k$-th RB.

According to\textcolor[rgb]{1.00,0.00,0.00}{\cite{b6}}, the total
power consumption $P({\bf{a}},{\bf{p}})$ for H-CRANs mainly depends
on the transmit power and circuit power. When the power consumption
for the fronthaul is considered, the total power consumption per RRH
is written as:
\begin{equation}\label{Eq.02}
P({\bf{a}},{\bf{p}}) = {\varphi _{{\rm{eff}}}}\sum\limits_{n =
1}^{N+M} {\sum\limits_{k = 1}^K {{a_{n,k}}{p_{n,k}}} }  +
P_{\rm{c}}^{\rm{R}} + {P_{{\rm{bh}}}},
\end{equation}
where $\varphi _{{\rm{eff}}}$, $P_{\rm{c}}^{\rm{R}}$ and
$P_{{\rm{bh}}}$ denote the efficiency of the power amplifier,
circuit power and power consumption of the fronthaul link,
respectively.

Similarly, the sum data rate for the HPN can be calculated as:
\begin{equation}\label{eqn:HPNrates}
C_M({\bf{a^M}},{\bf{p^M}}) = \sum\limits_{t = 1 }^{T}
{\sum\limits_{m = 1}^M {{a_{t,m}}{B_0}{{\log }_2}(1 + {\sigma
_{t,m}}{p_{t,m}})} },
\end{equation}
where $t \in \{1,...,T\}$ denotes the HUE allocated to the RB set
${\Omega _2}$. The $T \times M$ matrices ${\bf{a^M}} = {\left[
{{a_{t,m}}} \right]_{T \times M}}$ and ${\bf{p^M}} = {\left[
{{p^m_{t,m}}} \right]_{T \times M}}$ represent the feasible RB and
power allocation policies for the HPN, respectively. ${a_{t,m}}$ is
defined as the RB allocation indicator which can only be 1 or 0,
indicating whether the $m$-th RB is allocated to the HUE $t$.
${p^m_{t,m}}$ denotes the transmit power allocated to the HUE $t$ on
the $m$-th RB. ${\sigma _{t,m}}$ represents the CINR of the $t$-th
HUE on the $m$-th RB. The total power consumption of the HPN can be
given by
\begin{equation}\label{Eq.HPN_EE}
P_M({\bf{a^M}},{\bf{p^M}}) = {{\varphi}^M
_{{\rm{eff}}}}\sum\limits_{t = 1}^{T} {\sum\limits_{m = 1}^M
{{a_{t,m}}{p^M _{t,m}}} }  + P_{\rm{c}}^{\rm{M}} +
{P_{\rm{M_{bh}}}},
\end{equation}
where $p^M _{t,m}$ denotes the transmission baseband power for the
HPN when the $m$-th RB is allocated to the $t$-th HUE, which forms
the transmit power vector $\bf{p^M}$ for all HUEs.
${{\varphi}^M_{\rm{eff}}}$, $P_{\rm{c}}^{\rm{M}}$ and
$P_{\rm{M_{bh}}}$ denote the efficiency of power amplifier, the
circuit power and the power consumption of the backhaul link between
HPN and BBU pool, respectively. Considering that the HPN is mainly
utilized to extend the coverage and provide basic services for HUEs,
the same downlink transmit power for different HUEs over different
RBs is assumed as $p^M _{t,m} = P^M$, where $P^M$ has been defined
in \eqref{eqn:CINR1}.

The overall EE performance for the H-CRAN with \emph{L} RRHs and
\emph{1} HPN can be written as

\begin{eqnarray}
\gamma = \frac{{L * C({\bf{a}},{\bf{p}}) +
C_M({\bf{a^M}},{\bf{p^M}})}}{{L * P({\bf{a}},{\bf{p}}) +
P_M({\bf{a^M}},{\bf{p^M}})}}. \label{eqn:EE1}
\end{eqnarray}

For the dense RRH deployed H-CRAN, \emph{L} is much larger than 1.
The inter-tier interference from RRHs to HUEs remains constant when
the density of RRHs is sufficiently high, and hence the downlink SE
and EE performances for the HPN can be assumed to be stable.
Therefore, $C({\bf{a}},{\bf{p}})$ is much larger than
$C_M({\bf{a^M}},{\bf{p^M}})$, and $P_M({\bf{a^M}},{\bf{p^M}})$ can
be ignored if \emph{L* P({\bf{a}},{\bf{p}})} is sufficiently large.
When \emph{L} is sufficiently large, the overall EE performance in
\eqref{eqn:EE1} for the H-CRAN can be approximated as:

\begin{eqnarray}\label{eqn:EE_App}
\gamma \approx \frac{{L * C({\bf{a}},{\bf{p}})}}{{L *
P({\bf{a}},{\bf{p}})}} = \frac{{C({\bf{a}},{\bf{p}})}}{{
P({\bf{a}},{\bf{p}})}}. \label{eqn:EE2}
\end{eqnarray}

According to \eqref{eqn:EE2}, the overall EE performance mainly
depends on the EE optimization of each RRH. To make sure that HUEs
meet the low rate-constrained QoS requirement, the interference from
RRHs should be constrained and not larger than the predefined
threshold $\delta_0$. Consequently, to optimize downlink EE
performances, the core problem is converted to optimize EE
performance of each RRH with constraints on the inter-tier
interference to the HPN from RRHs when the density of RRHs is
sufficiently high.

\begin{prob}[\textbf{Energy Efficiency Optimization}] With the
constraints on the required QoS, inter-tier interference and maximum
transmit power allowance, the EE maximization problem in the
downlink H-CRAN can be formulated as
\begin{eqnarray}
\mathop {{\rm{max}}}\limits_{\{ {\bf{a}},{\bf{p}}\} }
&&\frac{{C({\bf{a}},{\bf{p}})}}{{P({\bf{a}},{\bf{p}})}} =
\frac{{\sum\limits_{n = 1}^{N + M} {\sum\limits_{k = 1}^K
{{a_{n,k}}{B_0}{{\log }_2}(1 + {\sigma _{n,k}}{p_{n,k}})} }
}}{{{\varphi _{{\rm{eff}}}}\sum\limits_{n = 1}^{N + M}
{\sum\limits_{k = 1}^K {{a_{n,k}}{p_{n,k}}} }  + P_{\rm{c}}^{\rm{R}}
+ {P_{{\rm{bh}}}}}}\\ \label{eqn:EEOP}
{\rm{s.t.}}&&\sum\limits_{n = 1}^{N + M} {{a_{n,k}}}  = 1, {a_{n,k}} \in \{ 0, 1 \},  \forall k,\label{eqn:c1}\\
&&\sum\limits_{k = 1}^K {C_{n,k}}  \ge {\eta _{\rm{R}}},  1 \le n \le N,\label{eqn:c2}\\
&&\sum\limits_{k = 1}^K {C_{n,k}}  \ge {\eta _{\rm{ER}}},  N + 1 \le n \le N + M,\label{eqn:c3}\\
&&\sum\limits_{n = N}^{N + M} {{a_{n,k}}{p_{n,k}}d_k^{{\rm{R2M}}}h_{k}^{{\rm{R2M}}}}  \le {\delta _0},  k \in {\Omega _{\rm{II}}},\label{eqn:c4}\\
&&\sum\limits_{n = 1}^{N + M} {\sum\limits_{k = 1}^K
{{a_{n,k}}{p_{n,k}}} }  \le P_{\max }^{\rm{R}}, {p_{n,k}} \ge
0,\forall k,\forall n,\label{eqn:c5}
\end{eqnarray}
where $C_{n,k}={a_{n,k}}{B_0}{{\log }_2}(1 + {\sigma _{n,k}}{p_{n,k}})$ and the constraint \eqref{eqn:c1} denotes the RB allocation
limitation that each RB cannot be allocated to more than one RUE at
the same time. The constraints of \eqref{eqn:c2} and \eqref{eqn:c3}
corresponding to the high and low rate-constrained QoS requirements
specify the minimum data rate of $\eta _{\rm{R}}$ and $\eta
_{\rm{ER}}$, respectively. \eqref{eqn:c4} puts a limitation on
${p_{n,k}}$ to suppress the inter-tier interference from RRHs to
HUEs that reuse the RB $k \in {\Omega _{\rm{II}}}$.
$d_k^{{\rm{R2M}}}$ and $h_k^{{\rm{R2M}}}$ represent the
corresponding path loss and channel gain on the $k$-th RB from the
reference RRH to the interfering HUE, respectively. In
\eqref{eqn:c5}, $P_{\max }^{\rm{R}}$ denotes the maximum transmit
power of the RRH.
\end{prob}

Based on the enhanced S-FFR scheme, the interference to HUE is
constrained, and the SINR threshold $\eta_{HUE}$, which is the
minimum SINR requirement for decoding the signal of HUE
successfully, is defined to represent the constraint of
\eqref{eqn:c4}. When allocating the \emph{m}-th RB to the
\emph{t}-th HUE, the SINR ($\eta_{t,m}$) larger than $\eta_{HUE}$
can be given by
\begin{equation}\label{eqn:SINR_HUE}
\eta_{t,m} = P_{}^{\rm{M}}d_m^{\rm{M}}h_{t,m}^{\rm{M}} /(L * \delta
_0 + {B_0}{N_0}) \geq {\eta_{HUE}}.
\end{equation}

Obviously, the optimal RB allocation policy $\bf{a^*}$ and power
allocation policy $\bf{p^*}$ with constraints of diverse QoS
requirements and variable $\eta_{HUE}$ to maximize the EE
performance in \emph{Problem 1} is a non-convex optimization problem
due to forms of the objective function and the RB allocation
constraint in \eqref{eqn:c1}, whose computing complexity increases
exponentially with the number of binary
variables\textcolor[rgb]{1.00,0.00,0.00}{\cite{b14}}. Intuitively,
\emph{Problem 1} is a mixed integer programming problem and the
fractional objective make it complicated and difficult to be solved
directly with the classical convex optimization methods.

\section{Energy-Efficient Resource Allocation Optimization}

In this section, we propose an effective method to solve
\emph{Problem 1} in (8), where we first exploit the non-linear
fractional programming for converting the objective function in
\emph{Problem 1}, upon which we then develop an efficient iterative
algorithm to solve this EE performance maximization problem.

\subsection{Optimization Problem Reformulation}

Since the objective function in \emph{Problem 1} is classified as a
non-linear fractional
programm\textcolor[rgb]{1.00,0.00,0.00}{\cite{b15}}, the EE
performance of the reference RRH can be defined as a non-negative
variable $\gamma$ in \eqref{eqn:EE_App} with the optimal value
$\gamma_{}^{*} =
\frac{C({\bf{a}_{}^{*}},{\bf{p}_{}^{*}})}{P({\bf{a}_{}^{*}},{\bf{p}_{}^{*}})}$.

\textbf{Theorem 1 (Problem Equivalence):} \emph{$\gamma^*$ is
achieved if and only if}
\begin{equation}
\mathop {{\rm{max}}}\limits_{\{{\bf{a}},{\bf{p}}\}}~
{C({\bf{a}},{\bf{p}})}-\gamma_{}^{*} {P({\bf{a}},{\bf{p}})}=
{C({\bf{a}_{}^{*}},{\bf{p}_{}^{*}})}-\gamma_{}^{*}
{P({\bf{a}_{}^{*}},{\bf{p}_{}^{*}})}=0,
\end{equation}
\emph{where $\{{\bf{a}},{\bf{p}}\}$ is any feasible solution of
Problem 1 to satisfy the constraints \eqref{eqn:c1}-\eqref{eqn:c5}.
\hfill\rule{8pt}{8pt}}

\textbf{\emph{Proof:}} Please see Appendix A. \hfill\rule{8pt}{8pt}

Based on the optimal condition stated in \textbf{Theorem 1},
\emph{Problem 1} is equivalent to \emph{Problem 2} as follows if we
can find the optimal value $\gamma^*$. Although $\gamma^*$ cannot be
obtained directly, an iterative algorithm (\textbf{Algorithm 1}) is
proposed to update $\gamma$ while ensuring that the corresponding
solution $\{{\bf{a}}, {\bf{p}}\}$ remains feasible in each
iteration. The convergence can be proved and the optimal RA to solve
\emph{Problem 2} can be derived.

\emph{Problem 2 (\textbf{Transformed Energy Efficiency
Optimization}): }
\begin{eqnarray}\label{eqn:OP2}
&\mathop {{\rm{max}}}\limits_{\{{\bf{a}},{\bf{p}}\}} & {C({\bf{a}},{\bf{p}})} - \gamma^* {P({\bf{a}},{\bf{p}})},\\
&{\text{s.t.}} & \eqref{eqn:c1}-\eqref{eqn:c5}.\nonumber
\end{eqnarray}

Note that \emph{Problem 2} is a tractable feasibility problem.
\hfill\rule{8pt}{8pt}

Hence, the objective function of the fractional form in (8) is
transformed into the subtractive form. To design the efficient
algorithm for solving \emph{Problem 2}, we can define an equivalent
function $F(\gamma)=\mathop
{{\rm{max}}}\limits_{\{{\bf{a}},{\bf{p}}\}} ~ {C({\bf{a}},{\bf{p}})}
- \gamma {P({\bf{a}},{\bf{p}})}$ with the following lemma.

\textbf{Lemma 1:} For all feasible $\bf{a}$, $\bf{p}$ and $\gamma$,
\emph{$F(\gamma )$ is a strictly monotonic decreasing function in
$\gamma$, and $F(\gamma ) \ge 0$ . \hfill\rule{8pt}{8pt}}

\textbf{\emph{Proof:}} Please see Appendix B. \hfill\rule{8pt}{8pt}

Due to the constraint of \eqref{eqn:c1}, the feasible domain of
$\bf{a}$ is a discrete and finite set consisting of all possible RB
allocations, and thus $F(\gamma )$ is generally a continuous but
non-differentiable function with respect to $\gamma$.

\subsection{Proposed Iterative Algorithm}

Based on \textbf{Lemma 1}, an iterative algorithm is proposed to
solve the transformed \emph{Problem 2} by updating $\gamma$ in each
iteration as the following \textbf{Algorithm 1}.

\begin{algorithm}[htb]
\renewcommand{\algorithmicrequire}{\textbf{Input:}}
\renewcommand\algorithmicensure {\textbf{Output:} }
\caption{Energy-Efficient Resource Assignment and Power Allocation}
\label{alg:Framwork}
\begin{algorithmic}[1]
\STATE Set the maximum number of iterations ${I_{\max}}$,
convergence condition ${\varepsilon _\gamma }$ and the initial value
${\gamma ^{(1)}} = 0$.\\
\STATE Set the iteration index $i=1$ and begin the iteration (Outer Loop).\\
\STATE \textbf{for} $1 \le i \le {I_{\max }}$ \\
\STATE ~~Solve the resource allocation problem with ${\gamma ^{(i)}}$ (Inner Loop); \\
\STATE ~~Obtain ${\bf{a}}^{(i)},{\bf{p}}^{(i)}$,
${C}({\bf{a}}^{(i)},{\bf{p}}^{(i)})$, ${P}({\bf{a}}^{(i)},{\bf{p}}^{(i)})$; \\
\STATE ~~\textbf{if} ${C}({\bf{a}}^{(i)},{\bf{p}}^{(i)})-{\gamma
^{(i)}}{P}({\bf{a}}^{(i)},{\bf{p}}^{(i)}) < {\varepsilon _\gamma }$ \textbf{then} \\
\STATE ~~~~Set $\{
{{\bf{a}}^*},{{\bf{p}}^*}\}\!\!=\!\!\{{{\bf{a}}^{(i)}},{{\bf{p}}^{(i)}}\}
$ and ${\gamma^*}\!\!=\!\!{\gamma^{(i)}}$; \\
\STATE ~~~~\textbf{break}; \\
\STATE ~~\textbf{else} \\
\STATE ~~~~Set ${\gamma ^{(i+1)}} =
\frac{{{C}({\bf{a}^{(i)}},{\bf{p}^{(i)}})}}{{{P}({\bf{a}^{(i)}},{\bf{p}^{(i)}})}}$
and $i=i+1$;
\STATE ~~\textbf{end if} \\
\STATE \textbf{end for}
\end{algorithmic}
\end{algorithm}

The proposed iterative \textbf{Algorithm 1} ensures that $\gamma$
increases in each iteration. It can be observed that two nested
loops executed in \textbf{Algorithm 1} can achieve the optimal
solution to maximize EE performances. The outer loop updates
$\gamma^{(i+1)}$ in each iteration with the ${C}({\bf{a}^{(i)}},
{\bf{p}^{(i)}})$ and ${P}({\bf{a}^{(i)}}, {\bf{p}^{(i)}})$ obtained
in the last iteration. In the inner loop, the optimal RB allocation
policy ${\bf{a}}^{(i)}$ and power allocation policy ${\bf{p}}^{(i)}$
with a given value of $\gamma^{(i)}$ are derived by solving the
following inner-loop \emph{Problem 3}.

\emph{Problem 3 \textbf{(Resource Allocation Optimization in the
Inner Loop)}: }
\begin{eqnarray}\label{eqn:OP3}
&\mathop {{\rm{max}}}\limits_{\{{\bf{a}},{\bf{p}}\}} & {C({\bf{a}},{\bf{p}})} - \gamma^{(i)} {P({\bf{a}},{\bf{p}})},\\
&{\text{s.t.}} & \eqref{eqn:c1}-\eqref{eqn:c5}, \nonumber
\end{eqnarray}
where $\gamma^{(i)}$ is updated by the last iteration in outer loop.
\hfill\rule{8pt}{8pt}

Actually, with the help of the proposed \textbf{Algorithm 1}, the
solution to \emph{Problem 3} is converged and the optimal solution
is presented. The global convergence has the following theorem.

\textbf{Theorem 2 (Global Convergence):} \emph{\textbf{Algorithm 1}
always converges to the global optimal solution of Problem 3}.
\hfill\rule{8pt}{8pt}

\textbf{\emph{Proof:}} Please see Appendix C. \hfill\rule{8pt}{8pt}

The optimization problem in \eqref{eqn:OP3} is non-convex and hard
to be solved directly. Generally speaking, if \eqref{eqn:OP3} can be
solved by the dual method, there exists a non-zero duality
gap\textcolor[rgb]{1.00,0.00,0.00}{\cite{b17}}. The duality gap is
defined as the difference between the optimal value of
\eqref{eqn:OP3} (denoted by $EE^*$) and the optimal value of the
dual problem for \eqref{eqn:OP3} (denoted by $D^*$). Fortunately, it
can be demonstrated that the duality gap between \eqref{eqn:OP3} and
the dual problem is nearly zero when the number of RBs is
sufficiently large\textcolor[rgb]{1.00,0.00,0.00}{\cite{b16}}, which
is illustrated as the following theorem.

\textbf{Theorem 3 (Duality Gap):} \emph{When the number of the
resource block is sufficiently large, the duality gap between
\eqref{eqn:OP3} and its dual problem is nearly zero, i.e., $D^*-EE^*
\approx 0$ holds}.\hfill\rule{8pt}{8pt}

\textbf{\emph{Proof:}} Please see Appendix D. \hfill\rule{8pt}{8pt}

\subsection{Lagrange Dual Decomposition Method}

Hence, according to \textbf{Theorem 3}, \emph{Problem 3} in the
$i$-th outer loop can be solved by the dual decomposition method.
With rearranging the constraints \eqref{eqn:c2}--\eqref{eqn:c5}, the
Lagrangian function of the primal objective function is given by
\begin{align}
& L({\bf{a}},{\bf{p}},{\bm{\beta }},{\bm{\lambda }},\nu ) = \sum\limits_{n = 1}^{N + M} {\sum\limits_{k = 1}^K {{a_{n,k}}{B_0}{{\log }_2}(1 + {\sigma _{n,k}}{p_{n,k}})} } \nonumber\\
&- \gamma^{(i)} \left( {{\varphi _{{\rm{eff}}}}\sum\limits_{n = 1}^{N + M} {\sum\limits_{k = 1}^K {{a_{n,k}}{p_{n,k}}} }  + P_{\rm{c}}^{\rm{R}} + {P_{{\rm{bh}}}}} \right) \nonumber \\
& + \sum\limits_{n = 1}^N {{\beta _n}\left[ {\sum\limits_{k = 1}^K {{a_{n,k}}{B_0}{{\log }_2}(1 \!+\! {\sigma _{n,k}}{p_{n,k}})}\! - \!{\eta _{\rm{R}}}} \right]} \! \nonumber\\
&+ \! \sum\limits_{n = N + 1}^{N + M} {{\beta _n}\left[ {\sum\limits_{k = 1}^K {{a_{n,k}}{B_0}{{\log }_2}(1 \!+\! {\sigma _{n,k}}{p_{n,k}})} \! -\! {\eta _{\rm{ER}}}} \right]} \nonumber \\
& + \sum\limits_{k = 1}^K {{\lambda _k}\left( {{\delta _0} -
\sum\limits_{n = 1}^{N + M}
{{a_{n,k}}{p_{n,k}}d_k^{{\rm{R2M}}}h_{k}^{{\rm{R2M}}}} } \right)}\nonumber\\
& +
\nu \left( {P_{\max }^{\rm{R}} - \sum\limits_{n = 1}^{N + M}
{\sum\limits_{k = 1}^K {{a_{n,k}}{p_{n,k}}} } } \right),
\end{align}
where $\bm{\beta}=(\beta_1, \beta_2, \ldots, \beta_{N+M})\succeq 0$
is the Lagrange multiplier vector associated with the required
minimum data rate constraints \eqref{eqn:c2} and \eqref{eqn:c3}.
$\bm{\lambda}=(\lambda_1, \lambda_2, \ldots, \lambda_{K})\succeq 0$
is the Lagrange multiplier vector corresponding to the inter-tier
interference constraint \eqref{eqn:c4} and $\lambda_k = 0$ for $k
\in {\Omega _{\rm{I}}}$. $\nu \ge 0$ is the Lagrange multiplier for
the total transmit power constraint \eqref{eqn:c5}. The operator
$\succeq 0$ indicates that the elements of the vector are all
nonnegative.

The Lagrangian dual function can be expressed as:
\begin{align} \label{eqn:LDU}
&g({\bm{\beta }},{\bm{\lambda }},\nu ) = \mathop {\max }\limits_{\left\{ {{\bf{a}},{\bf{p}}} \right\}} ~ L({\bf{a}},{\bf{p}},{\bm{\beta }},{\bm{\lambda }},\nu ) \nonumber \\
& = \mathop {\max }\limits_{\left\{ {{\bf{a}},{\bf{p}}} \right\}} ~ \Bigg\{ \sum\limits_{k = 1}^K {\sum\limits_{n = 1}^{N + M} {\Big[ {({\beta _n} + 1){a_{n,k}}{B_0}{{\log }_2}(1 + {\sigma _{n,k}}{p_{n,k}})} } } \nonumber\\
&- \gamma^{(i)} {\varphi _{{\rm{eff}}}}{a_{n,k}}{p_{n,k}} { - {\lambda _k}{a_{n,k}}{p_{n,k}}d_k^{{\rm{R2M}}}h_{k}^{{\rm{R2M}}} - \nu {a_{n,k}}{p_{n,k}}} \Big ]\nonumber  \\
&- \gamma^{(i)} (P_{\rm{c}}^{\rm{R}} + {P_{{\rm{bh}}}}) - \sum\limits_{n = 1}^N {{\beta _n}{\eta _{\rm{R}}}} \nonumber \\
& - \sum\limits_{n = N + 1}^{N + M} {{\beta _n}{\eta _{\rm{ER}}}}  +
\sum\limits_{k = 1}^K {{\lambda _k}{\delta _0}}  + \nu P_{\max
}^{\rm{R}} \Bigg\},
\end{align}
and the dual optimization problem is reformulated as:
\begin{eqnarray}
&\mathop {\min }\limits_{\left\{ {{\bm{\beta }},{\bm{\lambda }},\nu } \right\}} & g({\bm{\beta }},{\bm{\lambda }},\nu ), \\
&{\rm{s.t.}} & {\bm{\beta }} \succeq 0,{\bm{\lambda }} \succeq
0,\nu \ge 0. \nonumber
\end{eqnarray}

It is obvious that the dual optimization problem is always convex.
In particular, the Lagrangian function
$L({\bf{a}},{\bf{p}},{\bm{\beta }},{\bm{\lambda }},\nu )$ is linear
with ${\beta _n}$, ${\lambda _k}$ and $\nu$ for any fixed $a_{n,k}$
and $p_{n,k}$, while the dual function $g({\bm{\beta }},{\bm{\lambda
}},\nu )$ is the maximum of these linear functions. We use the dual
decomposition method to solve this dual problem, which is firstly
decomposed into $K$ independent problems as:

\begin{align} \label{eqn:df}
g({\bm{\beta }},{\bm{\lambda }},\nu ) &= \sum\limits_{k = 1}^K
{{g_k}({\bm{\beta }},{\bm{\lambda }},\nu )} \! -\! \gamma^{(i)}
(P_{\rm{c}}^{\rm{R}} + {P_{{\rm{bh}}}}) \! \nonumber \\
&- \! \sum\limits_{n =
1}^N {{\beta _n}{\eta _{\rm{R}}}} \!-\! \sum\limits_{n = N + 1}^{N +
M} {{\beta _n}{\eta _{\rm{ER}}}} +\! \sum\limits_{k = 1}^K {{\lambda
_k}{\delta _0}} +\! \nu P_{\max }^{\rm{R}},
\end{align}
where
\begin{align}\label{eqn:gk}
&{g_k}({\bm{\beta }},{\bm{\lambda }},\nu )\nonumber \\
 & = \mathop {\max }\limits_{\{ {\bf{a}},{\bf{p}}\} } ~ \Bigg \{ {\sum\limits_{n = 1}^{N + M} {\bigg [ {(1 + {\beta _n}){a_{n,k}}{B_0}{{\log }_2}(1 + {\sigma _{n,k}}{p_{n,k}})} } }\nonumber \\&
  - \gamma^{(i)} {\varphi _{{\rm{eff}}}}{a_{n,k}}{p_{n,k}}  { { - {\lambda
_k}{a_{n,k}}{p_{n,k}}d_k^{{\rm{R2M}}}h_{k}^{{\rm{R2M}}} - \nu
{a_{n,k}}{p_{n,k}}} \bigg]} \Bigg\}.
\end{align}

Supposed that the $k$-th RB is allocated to the $n$-th UE, i.e.,
$a_{n,k}=1$, it is obvious that \eqref{eqn:gk} is concave in terms
of ${p_{n,k}}$. With the Karush-Kuhn-Tucker (KKT) condition, the
optimal power allocation is derived by
\begin{equation} \label{eqn:power}
p_{n,k}^* = {\left[ {\omega _{n,k}^* - \frac{1}{{{\sigma _{n,k}}}}}
\right]^ + },
\end{equation}
where $\left[ x \right]^+ = \max \{x, 0 \}$, and the optimal
water-filling level $\omega _{n,k}^*$ is derived as
\begin{equation}\label{eqn:power_level}
\omega _{n,k}^* = \frac{{{B_0}(1 + {\beta_n})}}{{\ln
2(\gamma^{(i)}{\varphi _{{\rm{eff}}}}  + {\lambda
_k}d_k^{{\rm{R2M}}}h_{k}^{{\rm{R2M}}} + \nu )}}.
\end{equation}

Then, substituting the optimal power allocation obtained by
\eqref{eqn:power} into the decomposed optimization problem
\eqref{eqn:gk}, we can have
\begin{align}\label{eqn:gk_new}
&{g_k}({\bm{\beta }},{\bm{\lambda }},\nu )  \nonumber \\ &= \mathop
{\max }\limits_{1 \le n \le N+M} \Bigg\{ ( 1 \!+\! {\beta
_n}){B_0}{\left[ {{{\log }_2}(\omega _{n,k}^*{\sigma _{n,k}})}
\right]^ + }
 \!-\nonumber\\
 &\! (\gamma^{(i)}{\varphi _{{\rm{eff}}}} \!+\! {\lambda
_k}d_k^{{\rm{R2M}}}h_{k}^{{\rm{R2M}}}\!+\!\nu ){\left[ {\omega
_{n,k}^* \!\!-\!\! \frac{1}{{{\sigma _{n,k}}}}} \right]^ + }\Bigg\}.
\end{align}

With \eqref{eqn:power_level} and \eqref{eqn:gk_new}, the optimal RB
allocation indicator for the given dual variables can be expressed
as:

\begin{equation}
a_{n,k}^* = \left\{ {\begin{array}{*{20}{l}}
{1, ~{\rm{  }}n = \arg \mathop {\max }\limits_{1 \le n \le N+M} {H_{n,k}}},  \\
{0, ~{\rm{  otherwise,          }}}  \\
\end{array}} \right.
\end{equation}
where
\begin{align}
{H_{n,k}} &=  {\left[ {(1 + {\beta _n}){{\log }_2}(\omega
_{n,k}^*{\sigma _{n,k}})} \right]^ + }\nonumber\\
& - \frac{{(1 + {\beta _n})}}{{\ln 2}}{\left[ {1 - \frac{1}{{\omega
_{n,k}^*{\sigma _{n,k}}}}} \right]^ + }.
\end{align}

Then, the sub-gradient-based
method\textcolor[rgb]{1.00,0.00,0.00}{\cite{b19}} can be utilized to
solve the above dual problem and the sub-gradient of the dual
function can be written as
\begin{eqnarray}
&&\nabla \beta _n^{(l + 1)} = \sum\limits_{k = 1}^K {C_{n,k}^{(l)}}  - {\eta _{{\rm{R}}}},1 \le n \le N,\\
&&\nabla \beta _n^{(l + 1)} = \sum\limits_{k = 1}^K {C_{n,k}^{(l)}}  - {\eta _{{\rm{ER}}}},N + 1 \le n \le N + M,\\
&&\nabla \lambda _k^{(l + 1)} = \left\{ {\begin{array}{*{20}{c}}
0{,\forall k \in {\Omega _{\rm{I}}},}\\
{{\delta _0} - \sum\limits_{n = 1}^N
{a_{n,k}^{(l)}p_{n,k}^{(l)}d_k^{{\rm{R2M}}}h_{k}^{{\rm{R2M}}}}
}{,\forall k \in {\Omega _{{\rm{II}}}},}
\end{array}} \right.\\
&&\nabla {\nu ^{(l + 1)}} = {P_{\max }^{\rm{R}}} - \sum\limits_{n =
1}^N {\sum\limits_{k = 1}^K {a_{n,k}^{(l)}p_{n,k}^{(l)}} },
\end{eqnarray}
where $a_{n,k}^{(l)}$ and $p_{n,k}^{(l)}$ represent the RB
allocation and power allocation which are derived by the dual
variables of the $l$-th iteration, respectively and $C_{n,k}^{(l)}=a_{n,k}^{(l)}{B_0}{{\log }_2}(1 + {\sigma _{n,k}}p_{n,k}^{(l)})$. $\nabla \beta
_n^{(l + 1)}$, $\nabla \lambda _k^{(l + 1)}$ and $\nabla {\nu ^{(l +
1)}}$ denote the sub-gradient utilized in the $(l+1)$-th inner loop
iteration. Hence, the update equations for the dual variables in the
$(l+1)$-th iteration are given by
\begin{eqnarray}
&&\beta _n^{(l + 1)} = {\left[ {\beta _n^{(l)} - \xi _\beta ^{(l + 1)} \times \nabla \beta _n^{(l + 1)}} \right]^ + }{\rm{,}}\forall n, \\
&&\lambda _k^{(l + 1)} = {\left[ {\lambda _k^{(l)} - \xi _\lambda ^{(l + 1)} \times \nabla \lambda _k^{(l + 1)}} \right]^ + },\forall k, \\
&&{\nu ^{(l + 1)}} = {\left[ {{\nu ^{(l)}} - \xi _\nu ^{(l + 1)}
\times \nabla {\nu ^{(l + 1)}}} \right]^ + },
\end{eqnarray}
where $\xi _\beta ^{(l + 1)}$, $\xi _\lambda ^{(l + 1)}$ and $\xi _\nu
^{(l + 1)}$ are the positive step sizes.

\section{Results and Discussions}

In this section, the EE performance of the proposed H-CRAN and
corresponding optimization solutions are evaluated with simulations.
There are $N = 10$ RUEs located in each RRH with the high
rate-constrained QoS and allocated by the orthogonal RB set ${\Omega
_1}$. $M$ is varied to denote the number of RUEs with low
rate-constrained QoS requirements. In the case of $M = 0$, there are
no RUEs to share the same radio resources with HUEs, and thus there
are no inter-tier interferences. Otherwise, there are $M(
> 0)$ RUEs with the low rate-constrained QoS interfered by HPN. We assume that $d_n^{{\rm{P}}} = 50$ m, $d_n^{{\rm{M}}} = 450$ m in the case of $1 \le
n \le N$; and $d_n^{{\rm{P}}} = 75$ m, $d_n^{{\rm{M}}} = 375$ m in
the case of $N+1 \le n \le N+M$. The distance between the reference
RRH and HUEs who reuse the $k$-th RB is $d_k^{{\rm{R2M}}} = 125$ m.
The number of total RBs is $K = 25$ and the system bandwidth is 5
MHz. The total transmit power of HPN is 43 dBm and equally allocated
on all RBs. It is assumed that the path loss model is expressed as
$31.5 + 40.0*\log_{10}(d)$ for the RRH-to-RUE link, and $31.5 +
35.0*\log_{10}(d)$ for the HPN-to-RUE and RRH-to-HUE links, where
$d$ denotes the distance between transmitter and receiver in meters.
The number of simulation snapshots is set at 1000. In all snapshots,
the fast-fading coefficients are all generated as independently and
identically distributed (i.i.d.) Rayleigh random variables with unit
variances. The low and high rate-constrained QoS requirements are
assumed to be $\eta _{\rm{ER}} = 64$ kbit/s and $\eta _{\rm{R}} =
128$ kbit/s, respectively.

We assume the static circuit power consumption to be
$P_{\rm{c}}^{\rm{R}} = 0.1$ W, and the power efficiency to be
$\varphi _{{\rm{eff}}} = 2$ for the power amplifiers of RRHs. For
HPNs, they as assumed to be $P_{\rm{c}}^{\rm{M}} = 10.0$ W and
$\varphi _{{\rm{eff}}}^{\rm{M}} = 4$, respectively
\textcolor[rgb]{1.00,0.00,0.00}{\cite{b20}}. The power consumptions
for both RRHs and HPNs to receive the channel state indication (CSI)
and coordination signalling are taken into account, i.e., the power
consumptions of the fronthaul link $P_{\rm{bh}}$, and the backhaul
link between the HPN and the BBU pool $P_{\rm{bh}}$ are assumed to
be 0.2 W.

\subsection{H-CRAN Performance Comparisons}

To evaluate EE performance gains of H-CRANs, the traditional C-RAN,
2-tier HetNet, and 1-tier HPN scenarios are presented as the
baselines. Only one HPN and one RRH are considered in the H-CRAN
scenario. Similarly, only two HPNs are considered in the 1-tier HPN
scenario, and one HPN and one Pico base station (PBS) are considered
in the 2-tier HetNet scenario. For the 1-tier C-RAN scenario, two
RRHs are considered, in which one RRH is used to replace the HPN in
H-CRANs.

\begin{itemize}

\item[-] 1-tier HPN: All UEs access the HPN, and the RB set ${\Omega _1}$ and ${\Omega _2}$ are allocated to the cell-center-zone and cell-edge-zone UEs, respectively.
The optimal power and RB allocations are based on the classical
water-filling and the maximum signal-to-interference-plus-noise
ratio (SINR) scheduling algorithms, respectively. The number of
cell-center-zone UEs is 10, and the number of cell-edge-zone UEs
with the low rate-constrained QoS requirement is varied from 1 to
10.

\item[-] 2-tier overlaid HetNet: The orthogonal RB set ${\Omega _1}$ and ${\Omega _2}$ are allocated to the PBS and HPN,
respectively. The static circuit power consumption for the PBS is
$P_{\rm{c}}^{\rm{P}} = 6.8$ W, and the power efficiency is $\varphi
_{{\rm{eff}}}^{\rm{P}} = 4$ for the power amplifiers in PBS. Note
that RBs allocated to the PBS and HPN are orthogonal with no
inter-tier interferences. The number of RUEs served by the PBS keeps
constant and is set at 10, and the number of HUEs served by HPN is
varied. Note that UEs served by the HPN in this scenario are denoted
by HUEs with low rate-constrained QoS requirements to be consistent
with other baselines.

\item[-] 2-tier underlaid HetNet: All RBs in the set ${\Omega _1}$ and ${\Omega _2}$ are fully shared by UEs within the covering areas of the PBS and
HPN. The optimal power and RB allocations are adopted to enhance EE
performances\textcolor[rgb]{1.00,0.00,0.00}{\cite{bib:06}}. The
locations and number of UEs served by the HPN and PBS are same as
those in the 2-tier overlaid HetNet scenario.

\item[-] 1-tier C-RAN: All RBs in the set ${\Omega _1}$ and ${\Omega _2}$ are fully shared by RUEs. There are 10 cell-center-zone RUEs, and $M$ cell-edge-zone RUEs in each
RRH. Each RRH covers the same area coverage of the PBS in the 2-tier
HetNet scenario. The optimal power and RB allocations are based on
the classical water-filling and maximum SINR algorithms,
respectively.

\item[-] 2-tier H-CRAN: The orthogonal RBs in set ${\Omega _1}$ are only allocated to RUEs, while RBs in ${\Omega _2}$ are shared by cell-edge-zone RUEs and
HUEs, where the proposed optimal solution subject to the
interference mitigation and RRH/HPN association is used. The number
of cell-center-zone RUEs keeps constant and is set at 10, and the
number of cell-edge-zone RUEs is varied to evaluate EE performances.

\end{itemize}

\begin{figure}
\centering  \vspace*{0pt}
\includegraphics[width=0.4\textwidth]{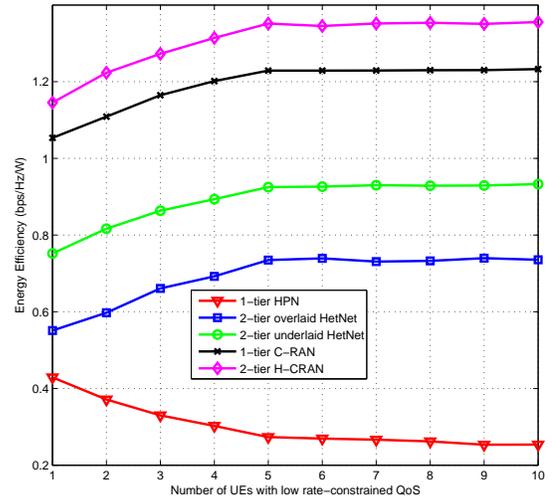}
\setlength{\belowcaptionskip}{-100pt} \caption{\textbf{Performance
comparisons among 1-tier HPN, 2-tier HetNet, 1-tier C-RAN, and
2-tier H-CRAN}}\label{HSRANgain} \vspace*{-10pt}
\end{figure}

As shown in Fig. \ref{HSRANgain}, EE performances are compared
amongst the 1-tier HPN, 2-tier underlaid HetNet, 2-tier overlaid
HetNet, 1-tier C-RAN, and 2-tier H-CRAN. The EE performance
decreases with the number of accessing UEs in the 1-tier HPN
scenario because more powers are consumed to make cell-edge-zone UEs
meet with the low rate-constrained QoS requirements. The EE
performance becomes better in the 2-tier HetNet than that in the
1-tier HPN because lower transmit power is needed and higher
transmission bit rate is achieved. Due to the spectrum reuse and the
interference is alleviated by the optimal solution
in\textcolor[rgb]{1.00,0.00,0.00}{\cite{bib:06}}, the EE performance
in the 2-tier underlaid HetNet scenario is better than that in the
2-tier overlaid HetNet scenario. Furthermore, the EE performance in
the 2-tier H-CRAN scenario is the best due to the gains from both
the H-CRAN architecture and the corresponding optimal radio resource
allocation solution. Note than the EE performance of 1-tier C-RAN is
a little worse than that of H-CRAN because it suffers from the
coverage limitation when RRH covers the area that the HPN serves in
the 2-tier H-CRAN. Meanwhile, the EE performance of 1-tier C-RAN is
better than that of both 1-tier HPN and 2-tire HetNet due to the
advantages of centralized cooperative processing and energy
consumption saving.

\subsection{Convergence of the Proposed Iterative Algorithm}

To evaluate EE performances of the proposed optimal resource
allocation solution (denoted by ``optimal EE solution''), two
algorithms are presented as the baselines. The first baseline
algorithm is based on the fixed power allocation (denoted by ``fixed
power"), i.e., the same and fixed transmit power is set for
different RBs, and the optimal power allocation derived in
\eqref{eqn:power} is not utilized. The second baseline algorithm is
based on the sequential RB allocation (denoted by ``sequential RB"),
where the RB is allocated to RUEs sequentially. The presented two
baseline algorithms use the same constraints of
\eqref{eqn:c1}-\eqref{eqn:c5} as the optimal EE solution does. There
are 12 RRHs (i.e., $L$ = 12) uniformly-spaced around the reference
HPN.

\begin{figure}
\centering  \vspace*{0pt}
\includegraphics[width=0.45\textwidth]{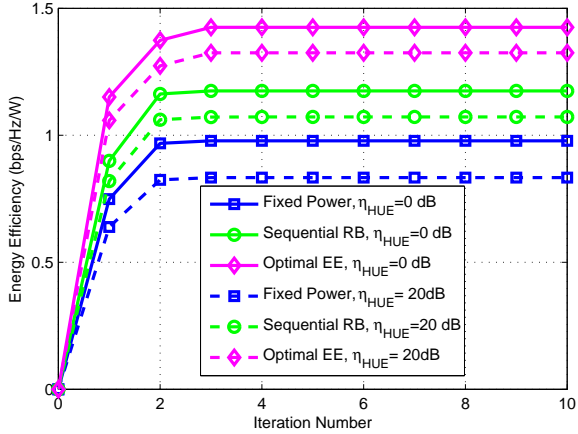}
\setlength{\belowcaptionskip}{-100pt} \caption{\textbf{Convergence
evolution under different optimization solutions}}\label{Converge}
\vspace*{-10pt}
\end{figure}

The EE performances of these three algorithms with different allowed
interference threshold under varied iteration numbers are
illustrated in Fig. \ref{Converge}. The number of RUEs with low
rate-constrained QoS requirement is assumed as $M = 3$. It can be
generally observed that the plotted EE performances are converged
within 3 iteration numbers for different optimization algorithms. On
the other hand, these two baseline algorithms are converged more
quickly than the proposal does because the proposal has a higher
computing complexity. Furthermore, the maximum allowed inter-tier
interference, which is defined as $\delta_0$ in \eqref{eqn:c4},
should be constrained to make the HUE work efficiently. To evaluate
$\delta_0$ impacting on the EE performance, $\eta_{HUE}$ described
in \eqref{eqn:SINR_HUE} denoting the maximum allowed inter-tier
interference is simulated. The large $\eta_{HUE}$ indicates that the
constraint of $\delta_0$ should be controlled to a low level, which
suggests to decrease the transmit power and even forbid the RB to be
allocated to RUEs. Consequently, the EE performance when
$\eta_{HUE}$ is 0 dB is better than that when it is 20 dB. The EE
performance decreases with the increasing SINR threshold
$\eta_{HUE}$ for these three algorithms.

\subsection{Energy Efficiency Performances of the Proposed Solutions}

In this part, key factors impacting on the EE performances of the
proposed radio resource allocation solution are evaluated. Since EE
performances are closely related to the constraints (12)--(13), the
variables $\eta_{HUE}$ and $P_{max}^R$ are two key factors to be
evaluated in Fig. 5 and Fig. 6, respectively. The ratio of
${\Omega_1}$ to ${\Omega_T}$ impacting on EE performances is
evaluated in Fig. 7 to show performance gains of the enhanced S-FFR.

In Fig. 5, EE performances under the varied SINR thresholds of HUEs
$\eta_{HUE}$ are compared among different algorithms when the
maximum allowed transmit power of RRH is 20 dBm or 30 dBm. When
$\eta_{HUE}$ is not sufficiently large, the EE performance almost
keeps stable with the increasing $\eta_{HUE}$ because the inter-tier
interference is not severe due to adoption of the enhanced S-FFR.
However, the EE performance declines for these two baseline
algorithms when the SINR threshold $\eta_{HUE}$ is over 10 dB. While
the proposed optimal EE solution can sustain more inter-tier
interferences, and the EE performance deteriorates after
$\eta_{HUE}$ is over 14 dB, indicating that the proposed solution
can mitigate more inter-tier interferences and provide higher bit
rates for HUEs than the other two baselines do. Summarily, the
proposed optimal EE solution can achieve the best EE performance,
while the ``fixed power'' algorithm has the worst EE performance,
and the ``sequential RB'' algorithm is in-between. Further, the EE
performances for these three algorithms are strictly related to the
maximum allowed transmit power of the RRH, and the EE performance
with $P_{max}^R$ = 30 dBm is better than that with $P_{max}^R$ = 20
dBm.

\begin{figure}
\centering  \vspace*{0pt}
\includegraphics[width=0.4\textwidth]{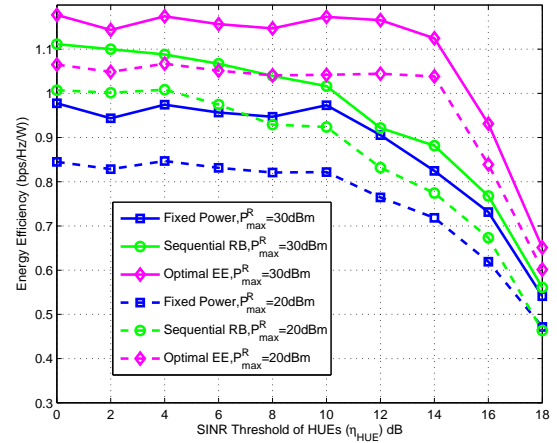}
\setlength{\belowcaptionskip}{-100pt} \caption{\textbf{EE
performance comparisons under different SINR thresholds of HUEs
$\eta_{HUE}$}}\label{SINRthreshold} \vspace*{-10pt}
\end{figure}

To further evaluate the impact of the maximum allowed transmit power
of RRHs $P_{max}^R$ on EE performances, Fig. 6 compares the EE
performance of different algorithms in terms of $P_{max}^R$. In this
simulation case, $\eta_{HUE}$ is set at 0 dB, the number of RUEs
with low rate-constrained QoS requirements is set at 4, and the
iteration number is set at 5. The maximum allowed transmit power of
the HPN is set at 43 dBm and the maximum allowed transmit power of
RRH varies within [14 dBm, 36 dBm] with the step size of 2 dBm. When
$P_{max}^R$ is not large, EE performances increase almost linearly
with the rising $P_{max}^R$ for all three algorithms. When
$P_{max}^R \geq 22 dBm$, both the SE performance and the total power
consumption increases almost linearly with the rising $P_{max}^R$.
Therefore, the EE performance almost keeps stable. As shown in Fig.
6, EE performances of the ``sequential RB" algorithm is often better
than those of the ''fixed power" algorithm due to the water-filling
power allocation gains indicated in Eq. (23). Further, the proposed
solution can achieve the best EE performance due to gains of RB
assignment and power allocation.

\begin{figure}
\centering  \vspace*{0pt}
\includegraphics[width=0.4\textwidth]{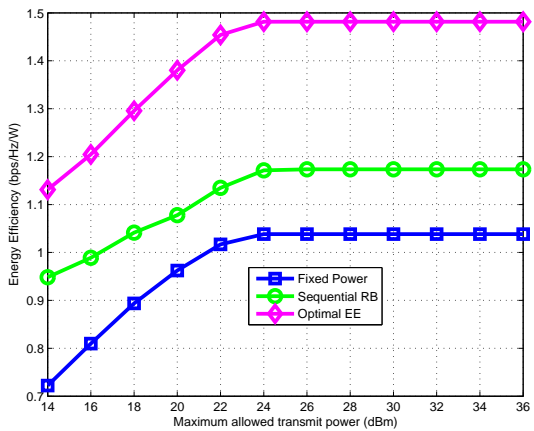}
\setlength{\belowcaptionskip}{-100pt} \caption{\textbf{EE
performance comparisons under different maximum allowed transmit
power of RRHs}}\label{txp} \vspace*{-10pt}
\end{figure}

Besides $\eta_{HUE}$ and $P_{max}^R$, the ratio between ${\Omega
_1}$ and ${\Omega _2}$ has a significant impact on the EE
performance of H-CRANs. The EE performances of the H-CRAN under the
enhanced S-FFR with different ratios of ${\Omega_1}$ to ${\Omega_T}$
are evaluated in Fig. 7, where the number of RUEs with low and high
rate constrained QoS requirements are set at $M = 5$ and $M = 10$,
respectively. Meanwhile, the number of total RBs is $K = 25$ and the
system bandwidth is 5 MHz. Each UE is allocated by at least one RB
to guarantee the minimal QoS requirement. Fig. 7 suggests that the
EE performance increases almost linearly with the ratio of ${\Omega
_1}$ to ${\Omega _T}$. With more available exclusive RBs for RRHs,
there are fewer inter-tier interferences because the shared RBs
specified for RUEs and HUEs become less. The increasing ratio of
${\Omega_1}$ to ${\Omega_T}$ results in the drastically increasing
of the expectation of SINRs for RUEs. Furthermore, since the
inter-tier interference is mitigated by configuring few shared RBs,
the transmission power of RUEs increases due to the proposed
water-filling algorithm in Eq. (23), which further improves overall
EE performances of H-CRANs. Besides, the EE performance increases
with the rising maximum allowed transmit power of RRH $P_{max}^R$,
which verifies the simulation results in Fig. 6 again. These results
indicate that more radio resources should be configured for ${\Omega
_1}$ if only the EE performance optimization is pursued. However,
the fairness of UEs should be considered jointly, and some necessary
RBs should be fixed for both HUEs and cell-edge-zone RUEs to
guarantee the seamless coverage and successful delivery of the
control signallings to all UEs in the real H-CRANs, which is a
challenging open issue for the future research.

\begin{figure}
\centering  \vspace*{0pt}
\includegraphics[width=0.5\textwidth]{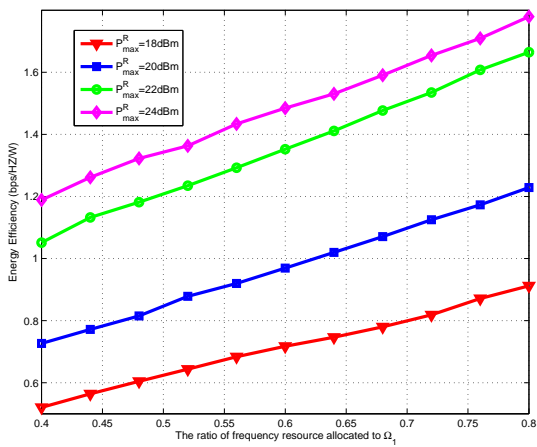}
\setlength{\belowcaptionskip}{-100pt} \caption{\textbf{EE
performance comparisons for different ratios of ${\Omega_1}$ to
${\Omega_T}$}} \label{rof}\vspace*{-10pt}
\end{figure}

\section{Conclusions}

In this paper, the energy efficiency performance optimization for
H-CRANs has been analyzed. In particular, the resource block
assignment and power allocation subject to the inter-tier
interference mitigation and the RRH/HPN association have been
jointly optimized. To deal with the optimization of resource
allocations, a non-convex fractional programming optimization
problem has been formulated, and the corresponding Lagrange dual
decomposition method has been proposed. Simulation results have
demonstrated that performance gains of H-CRANs over the traditional
HetNet and C-RAN are significant. Furthermore, the proposed optimal
energy-efficient resource allocation solution outperforms the other
two baseline algorithms. To maximize EE performances further, the
advanced S-FFR schemes with more RB sets should be researched, and
the corresponding optimal ratio of different RB sets should be
designed in the future.

\appendices

\section{Proof of \textbf{Theorem 1}}

By following a similar approach presented in \cite{bib:18}, we prove
the \textbf{Theorem 1} with two septated steps.

First, the sufficient condition of \textbf{Theorem 1} should be
proved. We define the maximal EE performance of \emph{Problem 1} as
$\gamma_{}^{*} =
\frac{C({\bf{a}_{}^{*}},{\bf{p}_{}^{*}})}{P({\bf{a}_{}^{*}},{\bf{p}_{}^{*}})}$,
where ${\bf{a}^*}$ and ${\bf{p}^*}$ are the optimal RB and power
allocation policies, respectively. It is obvious that
$\gamma_{}^{*}$ holds:

\begin{equation}
\gamma_{}^{*} =
\frac{C({\bf{a}_{}^{*}},{\bf{p}_{}^{*}})}{P({\bf{a}_{}^{*}},{\bf{p}_{}^{*}})}
\ge
\frac{C({\bf{a}_{}^{}},{\bf{p}_{}^{}})}{P({\bf{a}_{}^{}},{\bf{p}_{}^{}})},
\end{equation}
where $\bf{a}$ and $\bf{p}$ are the feasible RB and power allocation
policies for solving \emph{Problem 1}. Then, according to (35), we
can derive the following formuals:
\begin{equation}
\left\{ {\begin{array}{*{20}{l}}
{C({\bf{a}},{\bf{p}}) - \gamma _{}^*P({\bf{a}},{\bf{p}}) \le 0},\\
{C({\bf{a}}_{}^*,{\bf{p}}_{}^*) - \gamma
_{}^*P({\bf{a}}_{}^*,{\bf{p}}_{}^*) = 0}.
\end{array}} \right.
\end{equation}

Consequently, we can conclude that $\mathop
{{\rm{max}}}\limits_{\{{\bf{a}},{\bf{p}}\}}~
{C({\bf{a}},{\bf{p}})}-\gamma_{}^{*} {P({\bf{a}},{\bf{p}})} = 0$ and
it is achievable by the optimal resource allocation policies
${\bf{a}^*}$ and ${\bf{p}^*}$. The sufficient condition is proved.

Second, the necessary condition should be proved. Supposed that
${{\bf{\hat a}}^*}$ and ${{\bf{\hat p}}^*}$ are the optimal RB and
power allocation policies of the transformed objective function,
respectively, we can have ${C({\bf{\hat a}_{}^{*}},{\bf{\hat
p}_{}^{*}})}-\gamma_{}^{*}{P({\bf{\hat a}_{}^{*}},{\bf{\hat
p}_{}^{*}})}=0$. For any feasible RB and power allocation policies
$\bf{a}$ and $\bf{p}$, they can be expressed as:

\begin{equation}
{C({\bf{a}},{\bf{p}})}-\gamma_{}^{*} {P({\bf{a}},{\bf{p}})} \le
{C({\bf{\hat a}_{}^{*}},{\bf{\hat p}_{}^{*}})}-\gamma_{}^{*}
{P({\bf{\hat a}_{}^{*}},{\bf{\hat p}_{}^{*}})}=0.
\end{equation}

The above inequality can be derived as:
\begin{equation}
\frac{C({\bf{a}_{}^{}},{\bf{p}_{}^{}})}{P({\bf{a}_{}^{}},{\bf{p}_{}^{}})} \le \gamma_{}^{*}~{\rm{and}}~\frac{C({\bf{\hat a}_{}^{*}},{\bf{\hat p}_{}^{*}})}{P({\bf{\hat a}_{}^{*}},{\bf{\hat p}_{}^*})} = \gamma_{}^{*}.
\end{equation}

Therefore, the optimal resource allocation policies ${{\bf{\hat
a}}^*}$ and ${{\bf{\hat p}}^*}$ for the transformed objective
function are also the optimal ones for the original objective
function. The necessary condition of \textbf{Theorem 1} is proved.

\section{Proof of \textbf{Lemma 1}}

We can define an equivalent function as:
\begin{equation}
F(\gamma)=\mathop {{\rm{max}}}\limits_{\{{\bf{a}},{\bf{p}}\}}
{C({\bf{a}},{\bf{p}})} - \gamma {P({\bf{a}},{\bf{p}})},
\end{equation}
and we assume that $\gamma_{}^{1}$ and $\gamma_{}^{2}$ are the
optimal value for these two optimal RB allocation solution $\{
{{\bf{a}}_{}^{1}},{{\bf{p}}_{}^{1}} \}$ and $\{
{{\bf{a}}_{}^{2}},{{\bf{p}}_{}^{2}} \}$, where $\gamma_{}^{1} >
\gamma_{}^{2}$. Then,
\begin{eqnarray}
&&F({\gamma ^2}) = C({{\bf{a}}^2},{{\bf{p}}^2}) - {\gamma
^2}P({{\bf{a}}^2},{{\bf{p}}^2}) \nonumber \\ && >
C({{\bf{a}}^1},{{\bf{p}}^1}) - {\gamma
^2}P({{\bf{a}}^1},{{\bf{p}}^1})\nonumber\\
&&> C({{\bf{a}}^1},{{\bf{p}}^1}) -
{\gamma ^1}P({{\bf{a}}^1},{{\bf{p}}^1}) = F({\gamma ^1}).
\end{eqnarray}

Hence, $F(\gamma )$ is a strictly monotonic decreasing function in
terms of $\gamma$.

Meanwhile, let $\bf{a}'$ and $\bf{p}'$ be any feasible RB and power
allocation policies, respectively. Set $\gamma ' =
\frac{C({\bf{a}'},{\bf{p}'})}{P({\bf{a}'},{\bf{p}'})}$, then
\begin{eqnarray}
F(\gamma ')&&=\mathop {{\rm{max}}}\limits_{\{{\bf{a}},{\bf{p}}\}}
{C({\bf{a}},{\bf{p}})} - \gamma ' {P({\bf{a}},{\bf{p}})}\nonumber\\
&& \ge
{C({\bf{a}'},{\bf{p}'})} - \gamma ' {P({\bf{a}'},{\bf{p}'})} = 0.
\end{eqnarray}

Hence, $F(\gamma) \ge 0$.

\section{Proof of \textbf{Theorem 2}}

Supposed that $\gamma_{}^{(i)}$ and $\gamma_{}^{(i+1)}$ represent
the EE performance of the reference RRH in the $i$-th and $(i+1)$-th
iteration of the outer loop, respectively, where $\gamma_{}^{(i)} >
0$, and $\gamma_{}^{(i+1)} > 0$, neither of them is the optimal
value $\gamma^*$. Meanwhile, the $\gamma_{}^{(i+1)}$ is given by
${\gamma ^{(i + 1)}} =
\frac{{C({{\bf{a}}^{(i)}},{{\bf{p}}^{(i)}})}}{{P({{\bf{a}}^{(i)}},{{\bf{p}}^{(i)}})}}$,
where ${{\bf{a}}^{(i)}}$ and ${{\bf{p}}^{(i)}}$ are the optimal RB
and power solutions of \emph{Problem 3} in the $i$-th iteration of
the outer loop, respectively. Note that $\gamma^*$ is defined as the
achieved maximum EE performance for all feasible RA solutions
$\{{{\bf{a}}}, {{\bf{p}}}\}$, and thus $\gamma_{}^{(i+1)}$ cannot be
larger than $\gamma^*$, e.g., $\gamma_{}^{(i+1)} \le \gamma^*$. It
has been proved that $F(\gamma ) > 0$ in \textbf{Lemma 1} when
$\gamma$ is not the optimal valur to achieve the maximum EE
performance. Thus, the $F(\gamma_{}^{(i)})$ can be written as:

\begin{eqnarray}\label{eqn:F1}
&&F({\gamma ^{(i)}}) = C({{\bf{a}}^{(i)}},{{\bf{p}}^{(i)}}) - {\gamma ^{(i)}}P({{\bf{a}}^{(i)}},{{\bf{p}}^{(i)}}) \nonumber\\
&&= P({{\bf{a}}^{(i)}},{{\bf{p}}^{(i)}})({\gamma ^{(i + 1)}} -
{\gamma ^{(i)}}) > 0.
\end{eqnarray}

\eqref{eqn:F1} indicates that $\gamma ^{(i+1)} > \gamma ^{(i)}$ due
to $P({{\bf{a}}^{(i)}},{{\bf{p}}^{(i)}}) > 0$, which suggests that
$\gamma$ increases in each iteration of the outer loop in
\textbf{Algorithm 1}. According to \textbf{Lemma 1}, the value of
$F(\gamma )$ decreases with the increasing number of iterations due
to the incremental value of $\gamma$.

On the other hand, it has been proved that the optimal condition
$F(\gamma^* ) = 0$ holds in\textbf{ Theorem 1}. The
\textbf{Algorithm 1} ensures that $\gamma$ increases monotonically.
When the updated $\gamma$ increases to the achievable maximum value
$\gamma^*$, \emph{Problem 2} can be solved with $\gamma^*$ and
$F(\gamma^* ) = 0$. Then, the global optimal solutions
${{\bf{a}}^*}$ and ${{\bf{p}}^*}$ are derived. We update $\gamma$ in
the iterative algorithm to find the optimal value $\gamma^*$. It can
be demonstrated that $F(\gamma )$ converges to zero when the number
of iteration is sufficiently large and the optimal condition as
stated in \textbf{Theorem 1} is satisfied. Therefore, the global
convergence of \textbf{Algorithm 1} is proved.

\section{Proof of \textbf{Theorem 3}}

We can rewrite Eq. \eqref{eqn:OP3} as:
\begin{align}
&{C({\bf{a}},{\bf{p}})} - \gamma^{(i)} {P({\bf{a}},{\bf{p}})}\nonumber\\
&=\sum\limits_{n = 1}^{N + M} {\sum\limits_{k =1}^K
{{a_{n,k}}{B_0}{{\log }_2}(1 + {\sigma _{n,k}}{p_{n,k}})}
}\nonumber\\
&-\gamma^{(i)}({\varphi _{{\rm{eff}}}}\sum\limits_{n = 1}^{N+M}
{\sum\limits_{k = 1}^K {{a_{n,k}}{p_{n,k}}} }  +
P_{\rm{c}}^{\rm{R}} + {P_{{\rm{bh}}}})\nonumber\\
&=\sum_{k=1}^{K}\left ( {\sum_{n=1}^{N+M}{{a_{n,k}}{B_0}{{\log }_2}(1
+ {\sigma _{n,k}}{p_{n,k}})}}\right. \nonumber\\
&\left.{ - \sum_{n=1}^{N+M}\gamma^{(i)}\varphi
_{{\rm{eff}}}{a_{n,k}}{p_{n,k}}-\frac{\gamma^{(i)}}{K}\left (
P_{\rm{c}}^{\rm{R}} + {P_{{\rm{bh}}}} \right )}\right ).
\end{align}

It is obvious that for the given RB allocation scheme, if letting
\begin{align}
{f_k}\left( {{\bm{p}_{n,k}}}
\right)&=\sum_{n=1}^{N+M}{{a_{n,k}}{B_0}{{\log }_2}(1 + {\sigma
_{n,k}}{p_{n,k}})}\nonumber\\
&- \sum_{n=1}^{N+M}\gamma^{(i)}\varphi
_{{\rm{eff}}}{a_{n,k}}{p_{n,k}}-\frac{\gamma^{(i)}}{K}\left (
P_{\rm{c}}^{\rm{R}} + {P_{{\rm{bh}}}} \right ),
\end{align}
Eq. (43) can be written as ${C({\bf{a}},{\bf{p}})} - \gamma^{(i)}
{P({\bf{a}},{\bf{p}})} = \sum\limits_{k = 1}^K {{f_k}\left(
{{\bm{p}_{n,k}}} \right)}$, where $\bm{p}_{n,k}\in \mathbb{C}^{N}$,
and $f_{k}\left ( \cdot  \right ):\mathbb{C}^{N}\rightarrow
\mathbb{R}$ is not necessarily convex. Similarly, the constraints
(9)--(13) can be expressed as the function of $\bm{p}_{n,k}$ with
$\sum\limits_{k = 1}^K {{\bm{h}_k}\left( {{\bm{p}_{n,k}}} \right)}
\le \bm{0}$, where $\bm{h}_{k}\left ( \cdot \right
):\mathbb{C}^{N}\rightarrow \mathbb{R}^{L}$, and $L$ represents the
number of constraints. Thus, Eq. (17) can be expressed as:

\begin{eqnarray}
\label{eqn:EEOP1}{EE^* = \rm{max}}&&\sum\limits_{k = 1}^K {{f_k}\left( {{\bm{p}_{n,k}}} \right)} ,\\
{\rm{s.t.}}&&\sum\limits_{k = 1}^K {{\bm{h}_k}\left(
{{\bm{p}_{n,k}}} \right)}  \le \bm{0},
\end{eqnarray}
where $\bm{0}\in \mathbb{R}^{L}$. To prove the duality gap between
Eq. \eqref{eqn:OP3} and the optimal value of its dual problem $D^*$
is zero, a perturbation function $v\left ( \bm{H} \right )$ is
defined and can be written as:

\begin{eqnarray}
v\left ( \bm{H} \right )={\rm{max}}&&\sum\limits_{k = 1}^K {{f_k}\left( {{\bm{p}_{n,k}}} \right)} ,\\
{\rm{s.t.}}&&\sum\limits_{k = 1}^K {{\bm{h}_k}\left(
{{\bm{p}_{n,k}}} \right)}  \le \bm{H},
\end{eqnarray}
where $\bm{H}\in \mathbb{R}^{L}$ is the perturbation vector.

Following\textcolor[rgb]{1.00,0.00,0.00}{\cite{b17}}, if $v\left (
\bm{H} \right )$ is a concave function of $\bm{H}$, the duality gap
between $D^*$ and $EE^*$ is zero. Therefore, to prove the concavity
of $v\left ( \bm{H} \right )$, a time-sharing condition should be
demonstrated as follows.

\textbf{Definition 1 (Time-sharing Condition):} \emph{Let
$\bm{p}_{n,k}^{1*}$ and $\bm{p}_{n,k}^{2*}$ be the optimal solutions
of Eq. (47) to $v\left ( \bm{H}_{1} \right )$ and $v\left (
\bm{H}_{2} \right )$, respectively. Eq. (45) satisfies the
time-sharing condition if for any $v\left ( \bm{H}_{1} \right )$ and
$v\left ( \bm{H}_{2} \right )$, there always exists a solution
$\bm{p}_{n,k}^{3}$ to meet:}

\begin{eqnarray}
&\sum_{k=1}^{K}\bm{h}_{k}\left ( \bm{p}_{n,k}^{3} \right )\leq \alpha \bm{H}_{1}+\left ( 1-\alpha  \right )\bm{H}_{2},\\
&\sum_{k=1}^{K}f_{k}\left ( \bm{p}_{n,k}^{3} \right )\geq \alpha
f_{k}\left ( \bm{p}_{n,k}^{1*} \right )+\left ( 1-\alpha  \right
)f_{k}\left ( \bm{p}_{n,k}^{2*}\right ),
\end{eqnarray}
where $0\leq \alpha \leq 1$.

Now we should prove that $v\left ( \bm{H} \right )$ is a concave
function of $\bm{H}$. For some $\alpha$, it is easy to find
$\bm{H}_{3}$ satisfying $\bm{H}_{3}=\alpha \bm{H}_{1}+\left (
1-\alpha  \right )\bm{H}_{2}$. Let $\bm{p}_{n,k}^{1*}$,
$\bm{p}_{n,k}^{2*}$ and $\bm{p}_{n,k}^{3*}$ be the optimal solutions
with these constraints of $v\left ( \bm{H}_{1} \right )$, $v\left (
\bm{H}_{2} \right )$, and $v\left ( \bm{H}_{3} \right )$,
respectively. According to the definition of time-sharing condition,
there exists a $\bm{p}_{n,k}^{3}$ satisfying
$\sum_{k=1}^{K}\bm{h}_{k}\left ( \bm{p}_{n,k}^{3} \right )\leq
\alpha \bm{H}_{1}+\left ( 1-\alpha \right )\bm{H}_{2}$ and
$\sum_{k=1}^{K}f_{k}\left ( \bm{p}_{n,k}^{3} \right )\geq \alpha
f_{k}\left ( \bm{p}_{n,k}^{1*} \right )+\left ( 1-\alpha  \right
)f_{k}\left ( \bm{p}_{n,k}^{2*} \right )$. Since $\bm{p}_{n,k}^{3*}$
is the optimal solution to $v\left ( \bm{H}_{3} \right )$, it is
obvious that $\sum_{k=1}^{K}f_{k}\left ( \bm{p}_{n,k}^{3*} \right
)\geq \sum_{k=1}^{K}f_{k}\left ( \bm{p}_{n,k}^{3} \right )\geq
\alpha f_{k}\left ( \bm{p}_{n,k}^{1*} \right )+\left ( 1-\alpha
\right )f_{k}\left ( \bm{p}_{n,k}^{2*} \right )$ holds. Then, the
concavity of $v\left ( \bm{H} \right )$ is proved.

Since $v\left ( \bm{H} \right )$ is concave, Eq. \eqref{eqn:EEOP1}
could be proved to satisfy the time-sharing condition. The
time-sharing condition is always satisfied for the multi-carrier
system when the number of carriers goes to
infinity\textcolor[rgb]{1.00,0.00,0.00}{\cite{b16}}, such as the
OFDMA based H-CRAN system in this paper. We can let
$\bm{p}_{n,k}^{1}$ and $\bm{p}_{n,k}^{2}$ be two feasible power
allocation solutions. There are $\alpha$ percentages of the total
carriers to be allocated the power with $\bm{p}_{n,k}^{1}$, while
the remaining $\left ( 1-\alpha \right )$ percentages of the total
carriers are allocated the power with $\bm{p}_{n,k}^{2}$. Then
$\sum\limits_{k = 1}^K {{f_k}\left( {{\bm{p}_{n,k}}} \right)}$ is a
linear combination, which is expressed as $\alpha f_{k}\left (
\bm{p}_{n,k}^{1} \right )+\left ( 1-\alpha  \right )f_{k}\left (
\bm{p}_{n,k}^{2} \right )$. Therefore, the constraints are linear
combinations, and it is proved that Eq. \eqref{eqn:EEOP1} satisfies
the time-sharing condition. Consequently, $v\left ( \bm{H} \right )$
is a concave function of $\bm{H}$, and the duality gap between $D^*$
and $EE^*$ should be zero. This theorem is proved.

\end{document}